\documentclass[journal, final, 10pt]{IEEEtran}

\ifCLASSINFOpdf
\else
   \usepackage[dvips]{graphicx}
\fi
\usepackage{url}
\usepackage{xcolor}
\usepackage{lipsum}
\usepackage{graphicx}
\usepackage{adjustbox}
\usepackage{optidef}
\usepackage{amsmath, mathrsfs}
\usepackage{amsmath,amsfonts,amssymb,amsthm, bm}
\usepackage{mathtools, enumitem}
\usepackage{tikz}
\usepackage{tikz-3dplot}
\usepackage{comment}
\usetikzlibrary{shapes,decorations}
\usetikzlibrary{positioning}
\usepackage{tabularx}
\usepackage{hyperref}
\usepackage{float}
\usepackage{subcaption}
\usepackage{cite}
\usepackage{algorithmicx}
\usepackage{algorithm}
\usepackage{algpseudocode}
\usepackage{makecell, multirow}

\setcellgapes{3pt}

\newcommand\norm[1]{\left\lVert#1\right\rVert}
\DeclareMathAlphabet{\mathpzc}{OT1}{pzc}{m}{it}

\DeclareMathOperator*{\argmin}{arg\,min}
\newcommand\blfootnote[1]{%
  \begingroup
  \renewcommand\thefootnote{}\footnote{#1}%
  \addtocounter{footnote}{-1}%
  \endgroup
}

\newcommand{\G}{\mathcal{G}}
\newcommand{\M}{\mathcal{M}}

\newcommand{\Tr}{\textsf{T}}
\newcommand{\Tra}{\text{tr}}

\newcommand{\TV}{\mathpzc{V}}

\begin{document}

\title{Joint Data Inpainting and Graph Learning via Unrolled Neural Networks}
\makeatletter\def\Hy@Warning#1{}\makeatother
\author{
    {
        Subbareddy Batreddy\textsuperscript{\textsection}, Pushkal Mishra\textsuperscript{\textsection}, Yaswanth Kakarla, Aditya Siripuram \\
        \vspace{0.2cm}
        Indian Institute of Technology Hyderabad, India.
    }
}

\maketitle
\begingroup\renewcommand\thefootnote{
\begin{NoHyper}\textsection\end{NoHyper}}
\footnotetext{Equal contribution}
\begin{abstract}
Given partial measurements of a time-varying graph signal, we propose an algorithm to simultaneously estimate both the underlying graph topology and the missing measurements. The proposed algorithm operates by training an interpretable neural network, designed from the unrolling framework. The proposed technique can be used both as a graph learning and a graph signal reconstruction algorithm. This work enhances prior work in graph signal reconstruction by allowing the underlying graph to be unknown; and also builds on prior work in graph learning by tailoring the learned graph to the signal reconstruction task. 
\end{abstract}

\IEEEpeerreviewmaketitle

\section{Introduction}
Graphs are a natural way to represent a large class of irregularly structured signals obtained from numerous sources such as health monitoring devices \cite{graph2021sensors}, economic networks \cite{book2009economicnetworks}, meteorological stations† \cite{ortega2018graph}, transportation networks \cite{liu2021traffic}, and biological networks \cite{menoret2017neuroclassification, hu2015alzheimer}. The graph's vertices represent the signal components, and the edges encode the relation between various signal components. Graph signal processing (GSP) extends techniques and concepts from classical signal processing (e.g., the Fourier transform and frequencies) to such graph signals \cite{sandryhaila2014big} and by exploiting the information from the underlying graph, GSP aims to improve upon the traditional techniques \cite{ortega2018graph}. Extending the tools of classical signal processing to graph-based data is of interest to the signal processing community due to its proven success in applications such as graph filters \cite{liu2018filter}, sampling \cite{marques2015sampling, anis2014towards}, graph neural networks \cite{gama2018convolutional, gama2019stability}, and graph learning from data \cite{dong2018learning}.

In this work, we investigate the problems of \emph{graph-based data inpainting} and \emph{graph learning}. In real-world datasets, certain values may be absent due to various factors, which causes loss in data or including errors in data collection. % Data inpainting is the process of estimating the missing values using the available information. This method finds applications in various fields, including statistics, machine learning, and data analysis, particularly when handling incomplete or noisy data points in diverse types of datasets, such as time series data or sensor data.
Graph-based data inpainting is the process of reconstructing missing data points from datasets that are derived from an underlying graph structure. Examples of such datasets include neurological data like fMRI (where the underlying graph is the functional connectivity map of the brain) \cite{richiardi2013machine}, temperature data (the underlying graph corresponds to geographical similarity) \cite{dong2016learning} and neuroskeletal data (underlying graph is based on neural connectivity) \cite{ortega2018graph}. Recent works \cite{chen2021time, chen2015signal, qiu2017time} have proposed an inpainting algorithm for time-varying graph signals. The inpainting task is accomplished by assuming some prior on the signal, typically via graph-variation minimization: which requires knowledge of the underlying graph. However, in many real-world applications, the underlying graph is unknown; as evident in neurological and biological datasets (such as fMRI and MEG), sensor measurements, and others.

This naturally points us to the problem of graph learning: constructing an approximate graph given a dataset assumed to be derived from an underlying graph. Techniques from GSP have offered a new perspective on the problem of graph learning by assuming certain priors on the data model. For example, well-known methods, such as \cite{dong2016learning, kalofolias2016learn, chepuri2017learning}, assume that the data is smooth on the graph (or that the signal has low graph frequencies): similar to assumptions made for data inpainting. Other techniques exploit statistical properties \cite{lake2010discovering, egilmez2017graph} or spectral characteristics \cite{thanou2017learning, egilmez2018graph, maretic2017graph}. Review articles such as \cite{dong2019learning, mateos2019connecting} provide comprehensive overviews of graph learning methods based on these different perspectives and assumptions. While most of these methods are not specific to time-varying data, techniques in \cite{kalofolias2017learning, yamada2020time} give algorithms for learning graphs from time-varying data. % as well, and techniques in \cite{yamada2020time} learn time-varying graphs from data.
%Time-varying graph learning modeling shows how the graphs change over time while incorporating constraints on these temporal variations. Time-varying graphs represent data where the relationships between nodes or entities change over time \cite{kalofolias2017learning, yamada2020time}.  

We identify two key challenges in deploying these graph learning algorithms for data inpainting. First, note that graph learning techniques are typically evaluated on a GSP task (such as classification or data inpainting) using the learned graph \cite{humbert2021learning, qiu2017time, kroizer2019modeling}. However, graph learning techniques may not be tailored to the GSP task (data-inpainting) at hand; as this is an \emph{open-loop system}. Secondly, there is a multitude of signal priors to deploy for graph learning: take for instance the global smoothness-based techniques from \cite{dong2016learning} which assume the signals are smooth on the graph, or the data inpainting techniques in \cite{chen2021time, qiu2017time} where the temporal difference of the signal is assumed to be smooth on the graph. It may not be clear which model fits the given dataset. Our work aims to investigate a systematic solution to address these challenges.

Consider starting with a parameterized graph-learning model (which encompasses existing models as special cases) and a dataset with missing entries. Our objective is to jointly learn the graph structure and estimate the missing entries. We first perform data-inpainting using a guessed graph structure and then use a suitable graph learning technique to update the graph. If the estimation results are not satisfactory, we update the parameters based on the received feedback. This iterative process forms a \emph{closed loop system} as opposed to the open loop system discussed earlier; forming the core of our approach. The parameter updates are accomplished via the unrolling framework \cite{monga2021algorithm}: where each iteration in an iterative technique is interpreted as a layer of a deep neural network. End-to-end training of this neural network accomplishes the role of feedback. The block diagram as shown in Figure \ref{fig:feedback-block} provides an illustration of the proposed model.
\vspace{-0.15cm}

\begin{figure}[!h]
    \centering
    \includegraphics[width=3.4in]{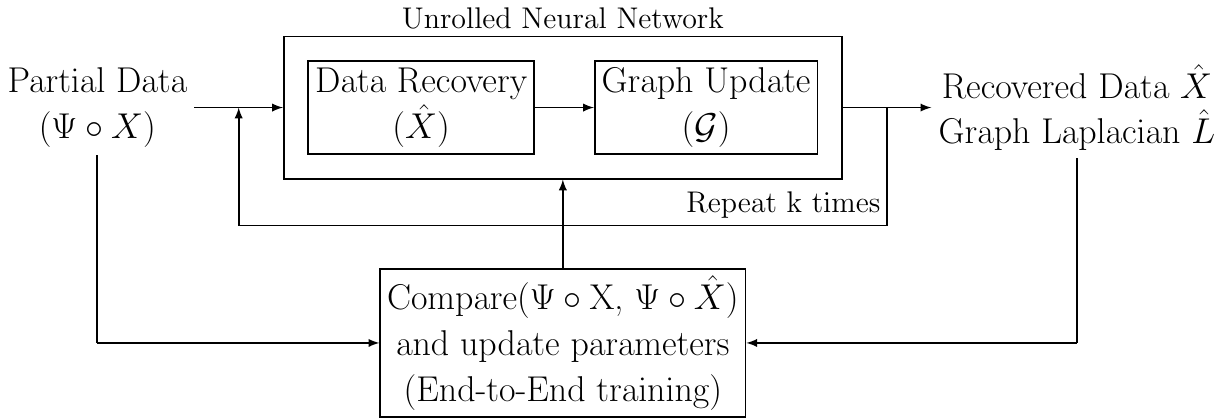}
    \caption{Proposed model with a closed-loop feedback system}
    \label{fig:feedback-block}
\end{figure}
\vspace{-0.15cm}

Thus the proposed approach \emph{jointly} accomplishes both data-inpainting and graph learning tasks. Note that this technique can also be used as a standalone graph-learning technique (by artificially removing a small number of entries from the given dataset).

Our work is heavily motivated from \cite{chen2021time}; where an unrolling framework is applied for data-inpainting with the underlying graph structure assumed to be known. We extend their work by allowing an unknown underlying graph. Consequently, the unrolled neural network in our setting is significantly different from the network in \cite{chen2021graph} (details in Section \ref{sec:unrolling}). Evaluation of our results show that: 
\begin{enumerate}
    \item The proposed algorithm estimates the missing entries better than techniques that first learn the graph and then use the learned graph for data inpainting (on real datasets)
    \item The proposed algorithm recovers a graph closer to the ground truth compared with other graph learning techniques (on synthetic datasets)
\end{enumerate}

Thus suggesting a potential use for the algorithm in both data-inpainting and graph-learning tasks. We hope that the proposed framework can also extend to GSP tasks other than inpainting as well, e.g. classification tasks; and allow for tailoring to other GSP tasks. %estimating relationships between companies based on historical stock price, identifying and constructing biological networks, and many more.
\section{Problem setup}
A data matrix $\mathbf{X}=[\Bar{x}_{1}, \Bar{x}_{2}, \dots , \Bar{x}_{M}]$ is an $N\times M$ matrix of real numbers, with each of the $M$ columns corresponding to signals at $M$ successive timestamps. We work with undirected graphs $\mathcal{G} = (V,E)$ with vertices $V=\{1,2,\ldots,N\}$ and edges $E\subseteq V \times V$. Each column $\Bar{x}_i: V\mapsto \mathbb{R}$ is interpreted as a graph signal defined on the vertices $V$ of $\G$. The Laplacian matrix $\mathbf{L}$ of the graph has a diagonal entry at $(i, i)$ as the degree of vertex $i$, an off-diagonal entry at $(i, j)$ as $-1$ if vertices $i$ and $j$ are connected by an edge, and $0$ otherwise. The graph-variation of the data matrix $\textbf{X}$ is defined as
\vspace{-0.1cm}
\begin{equation}
    \small
    \begin{aligned}\label{eq:graphsmoothness}
    \TV_{\G}(\textbf{X}) = Tr(\textbf{X}^{\Tr} \mathbf{L} \, \textbf{X}) 
    = \frac{1}{2} \sum_{t\;=\;1}^{M} \, \sum_{i \, \sim \, j} (x_{ti} - x_{tj})^{2}
    \end{aligned}
    \nonumber
\end{equation}
Graph-based data-inpainting algorithms \cite{chen2021time, qiu2017time} and graph learning algorithms \cite{dong2016learning, kalofolias2016learn, chepuri2017learning} assume that $\TV_{\G}(f(\textbf{X}))$ is small, for a suitably defined $f$, thus linking the data matrix with the graph structure. This assumption is then used to recover missing entries and the graph.

We denote $\Psi$ as the mask matrix where $\Psi_{ij} = 1$ if the entry at that location $(i, j)$ in $\mathbf{X}$ is known and $\Psi_{ij} = 0$ otherwise. With this setup, the problem under investigation is: `Given $\Psi \circ \textbf{X} + \textbf{E}$ (where $\textbf{E}$ is the noise) and $\Psi$, design an algorithm that outputs the graph $\G$ and the entire data matrix $\textbf{X}$; such that $\TV_{\G}(f(\textbf{X}))$ is small'. Some general notations: We denote by $\textbf{A} \circ \textbf{B}$ the Hadamard (elementwise) product of $\textbf{A}$ and $\textbf{B}$, by $\textbf{1}$ a vector of all $1'$s and tr(.) as the trace of the matrix. % We will use upper and lower case boldface letters to denote matrices and column vectors. 

\vspace{-0.25cm}
\section{Proposed algorithm}
\label{sec:proposed}
% \vspace{-0.1cm}
\subsection{Framing the optimization problem}
Motivated by \cite{chen2021time} and \cite{qiu2017time}, we define the function $f$ referenced in the previous section as a higher-order temporal difference. Let the temporal difference operator be defined as:
\vspace{-0.2cm}
\begin{equation}
    \small
    \begin{aligned}
    \label{eq:XDelta}
    \mathbf{X}\Delta = 
    \begin{bmatrix}
        \Bar{x}_{2} - \Bar{x}_{1} &
        \Bar{x}_{3} - \Bar{x}_{2} &
        \dots &
        \Bar{x}_{M} - \Bar{x}_{M-1}
    \end{bmatrix}
    \end{aligned}
    \nonumber
\end{equation}

Thus, we can take $f(\mathbf{X})=\mathbf{X}\Delta$, and get $\TV_{\G}(\mathbf{X}) = \text{Tr}(\mathbf{X}^\textsf{T} \mathbf{L} \mathbf{X}\Delta \Delta^\Tr)$. This variation measures the smoothness of the temporal difference on the graph. We generalize the $\Delta \Delta^\Tr$ operator above by introducing powers of $\Delta \Delta^\Tr$ and their polynomial combinations- this has the effect of accounting for smoothness of higher order temporal differences:
\vspace{-0.1cm}
\begin{equation}
    \small
    \label{eq:Z}
    Z(\Bar{\alpha}) = \alpha_0 \textbf{I} + \sum_{i=1}^{k} \alpha_i (\Delta \Delta^{\Tr})^{i}.
\end{equation}
\vspace{-0.2cm}

Hence, we use $\TV_{\G}(\mathbf{X}, \Bar{\alpha}) = \text{Tr}(\mathbf{X}^\textsf{T} \mathbf{L} \mathbf{X}Z(\Bar{\alpha}))$. Note that this is a semi-norm for $\Bar{\alpha}\geq 0$, and generalizes the regularizer (based on difference operators) from \cite{qiu2017time}. Now given the partial data $\Psi \circ \textbf{X}$ and $\Psi$, we frame the following optimization problem to obtain the complete data $\mathbf{\hat{X}}$ (output of the neural network), the graph Laplacian $\mathbf{L}$ (learned graph) and the model parameters $\bar{\alpha}$ (graph learning parameters):
\vspace{-0.4cm}

\begin{equation}
    \small
    \label{eq:overalleq}
    \begin{aligned}
    \min_{\hat{\mathbf{X}}, \mathbf{L}, \Bar{\alpha}} \norm{\Psi \circ (\mathbf{X} - \hat{\mathbf{X}})}^{2}_{F} + \lambda \TV_{\G}(\hat{\mathbf{X}}, \Bar{\alpha}) + \beta \norm{\mathbf{L}}^{2}_{F} + \gamma \norm{Z(\Bar{\alpha})}^{2}_{F}
    \\
    \text{s.t.} \quad \mathbf{L}_{ii} \geq 1, \mathbf{L}_{ij} = \mathbf{L}_{ji} \leq 0 \; \forall \; i\neq j, \mathbf{L} \cdot \mathbf{1}=\mathbf{0}
    \end{aligned}
\end{equation}

In the objective function, the first term ensures data fidelity, the second term is the graph variation, the third term ensures that the learned graph is sparse and the last term imposes a norm-bound constraint on the solution space and ensures stability of the solution. The constraints ensure that the learned $\mathbf{L}$ is a valid Laplacian matrix. % The above equation involves three sets of variables, the first set $\hat{\mathbf{X}}$ represents the output of neural network with estimated entries, the second set $\mathbf{L}$ represents the unknown graph to be learned, and the final set $\Bar{\alpha}$ represents the graph learning parameters. 
\vspace{-0.3cm}

\subsection{Solving the optimization problem}
The objective function in \eqref{eq:overalleq} is non-convex due to the product term involving the variables $\hat{\mathbf{X}}$, $\mathbf{L}$, and $\Bar{\alpha}$. To solve this issue, we adopt a standard alternating minimization technique. We fix the graph learning parameter $\Bar{\alpha}$ and alternately updating the Laplacian $\mathbf{L}$ and data matrix $\hat{\mathbf{X}}$ using projected gradient and conjugate gradient descent methods, respectively. Further, these steps are unrolled into a neural network to update the graph learning parameters $\Bar{\alpha}$.

\subsubsection{Estimating \texorpdfstring{$\hat{\mathbf{X}}$}{X}}
\label{sec:updatingX}
In the first step, we estimate unknown entries of the data matrix $\Psi \circ \mathbf{X}$, assuming $\mathbf{L}$ (initialized using covariance matrix from available data) and $\Bar{\alpha}$ are known. The corresponding optimization problem is as follows:

\vspace{-0.1cm}
\begin{equation}\label{eq:opstep1}
    \small
    \begin{aligned}
    \min_{\hat{\mathbf{X}}} \norm{\Psi \circ (\mathbf{X} - \hat{\mathbf{X}})}^{2}_{F} + \lambda \; \TV_{\G}(\hat{\mathbf{X}}, \Bar{\alpha})
    \end{aligned}
\end{equation}

This is solved using the conjugate gradient descent method \cite{boyd2004convex}, similar to the prior work in \cite{chen2021time}. Refer to appendix section \ref{sec:appendix} for more details (EMD: Estimate Missing Data).

\subsubsection{Updating \texorpdfstring{$\mathbf{L}$}{L}}
\label{sec:updatingL}
In the second step, we update \textbf{L} by fixing the other two variables $\hat{\mathbf{X}}$ and $\Bar{\alpha}$. The corresponding optimization problem is as follows:
\vspace{-0.3cm}

\begin{align*}
    \small
    \label{eq:opstep2}
    & \argmin_{\mathbf{L}} \quad \TV_{\G}(\hat{\mathbf{X}}, \Bar{\alpha}) + \beta \norm{\mathbf{L}}^{2}_{F}
    \\
    & \text{s.t. } \mathbf{L}_{ii} \geq 1, \mathbf{L}_{ij} = \mathbf{L}_{ji} \leq 0 \; \forall \; i\neq j, \mathbf{L} \cdot \mathbf{1}=\mathbf{0}
\end{align*}

To solve this optimization problem, we employ the Projected Gradient Descent method \cite{boyd2004convex}: this is a standard gradient descent followed by the projection of the updated variable onto a feasible space. The update steps are as follows:

\begin{algorithm}[ht]
    \renewcommand{\thealgorithm}{}
    \small
    \caption{\textbf{Graph Learning}}
    \label{sec:Graph_learning}
    \begin{algorithmic}[1]
        \Function{\texttt{GL~}}{$\Psi \circ \textbf{X}$, $\hat{\textbf{X}}$, $\mathbf{L}$, $\Bar{\alpha}$, $k$, $\beta$, $\eta$}
            % \State  \textbf{Initialization:} Covariance graph from $\Psi \; \circ$ \textbf{X}
            \State \textbf{Set:} $\mathbf{L}_1 = \mathbf{L}$ and $\M = \textbf{I} - {\textbf{1}} . {\textbf{1}}^{\Tr}$
            \For{$i \gets 1$ to $k$}
                \State Compute gradient $\triangledown f(\mathbf{L}_{i}) \leftarrow \hat{\mathbf{X}} \; Z(\Bar{\alpha}) \; \hat{\mathbf{X}}^{\Tr} + \beta \; \mathbf{L}_i$
                \State Update Laplacian $\mathbf{\Bar{L}}_{i+1} \leftarrow \mathbf{L}_{i} -\eta \triangledown f(\mathbf{L}_{i})$
                \State Project to feasible space $\mathbf{L}_{i+1} \leftarrow \text{Proj}_{\M}(\mathbf{\Bar{L}}_{i+1})$
            \EndFor
            \State \Return{$\mathbf{L}_{k}$}
        \EndFunction
    \end{algorithmic}
\end{algorithm}
% \vspace{-0.1cm}

Here we construct the $\text{Proj}_{\M}(\textbf{L})$ operator as: first extract the off-diagonal entries of $\textbf{L}$ and multiply them by $-1$ (hadamard product with $\M$), second passing them through a ReLU unit to retain edges with positive weights and finally converting resulting matrix into a Laplacian matrix by replacing the diagonal entries suitably. The update procedure for $\textbf{L}$ is better illustrated in Figure \ref{fig:proj-step}. %The use of non-linear projection onto the valid graph structure is crucial as it enables us to harness the power of neural networks.

\begin{figure}[ht]
    \centering
    \includegraphics[width=9cm]{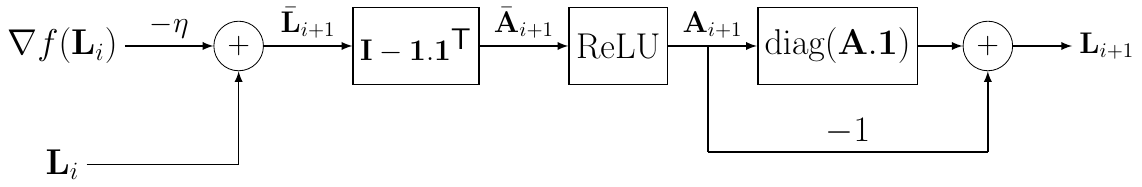}
    \caption{Illustration of the proposed Graph learning update step using linear operations and a ReLU unit}
    \label{fig:proj-step}
\end{figure}
\vspace{-0.8cm}

\subsection{Setting up the unrolled neural network}
\label{sec:unrolling}
Now, we have an iterative technique to find the graph and missing entries assuming $\Bar{\alpha}$ is fixed. Next, we unroll this algorithm by mapping each iteration to a single network layer. Our iterations consist of those that update update $\hat{\textbf{X}}$ (Section \ref{sec:updatingX}) and update \textbf{L} (Section \ref{sec:updatingL}). The design of our iterative technique for graph learning involves linear operations mixed with ReLU (Figure \ref{fig:proj-step}), which enables us to do the mapping easily. %The mapping for the rest of the data-inpainting steps to network layers is similar to \cite{chen2021time}.

The feed-forward process of this neural network is thus equivalent to iteratively reconstructing a graph and time-varying graph signal inpainting with the given parameters $\Bar{\alpha}$. Note that $\Psi \circ \textbf{X}$ is the input and $\hat{\textbf{X}}, \mathbf{L}$ are the outputs, and $\Bar{\alpha}$ the parameters of the neural network. We use the following loss function to train the neural network:
\vspace{-0.4cm}

\begin{equation}
    \small
    \begin{split}
        \label{eq:loss}
        \text{Loss}(\Bar{\alpha}) = \norm{\Psi \circ (\mathbf{X} - \hat{\mathbf{X}})}^{2}_{F} + \lambda \TV_{\G}(\hat{\mathbf{X}}, \Bar{\alpha}) \; + \beta \norm{\mathbf{L}}^{2}_{F} \\
        + \gamma \norm{Z(\Bar{\alpha})}^{2}_{F}
    \end{split}
\end{equation}
%Thus, we interlace both graph learning and data inpainting blocks to form an interpretable neural network. 
The previous discussion is summarized in the pseudocode below.
\begin{algorithm}[ht]
    \renewcommand{\thealgorithm}{}
    \small
    \caption{\textbf{Unrolled Neural Network (Forward pass)}}
    \label{sec:algorithm}
    \begin{algorithmic}[1]
        \State \textbf{Given:} Missing data matrix $\Psi \circ \mathbf{X}$ and $\Psi$
        \State \textbf{Input:} $\textbf{Y} = \Psi \circ \textbf{X}$ and $\Psi$
        \State \textbf{Hyperparameters:} $k$, $k_1$, $k_2$, $\eta$, $\beta$, $\gamma$ and $\lambda$
        \State \textbf{Parameters of the neural network:} $\Bar{\alpha}$
        \State  \textbf{Initialization:} $\mathbf{L}_1\leftarrow$ Covariance graph from $\Psi\circ \mathbf{X}$
        \For{$i \gets 1$ to $k$} \Comment{Unrolled iterations}
            \State $\hat{\mathbf{X}}_i \leftarrow \texttt{EMD~}(\textbf{Y}, \mathbf{L}_i, \Bar{\alpha}, \lambda, k_1)$ \Comment{See Appendix \ref{sec:appendix}}
            \State $\mathbf{L}_{i+1} \leftarrow $ {\texttt{GL~}}({$\textbf{Y}$, $\hat{\textbf{X}}_i$, $\mathbf{L}_{i}$, $\Bar{\alpha}$, $k_2$, $\beta$, $\eta$}) \Comment{Graph update}
        \EndFor 
        \State \textbf{Output} {$\mathbf{\hat{X}}, \mathbf{L}$}
    \end{algorithmic}
\end{algorithm}
% \vspace{-1cm}

\subsection{Differences from prior work}
Our work is motivated from and builds on \cite{chen2021time}. The main differences stem from the formulation of the objective function (\ref{eq:overalleq}): the graph is unknown for our framework, i.e. $\textbf{L}$ is a variable. In addition, we omit the polynomial terms in $\textbf{L}$ (to promote convexity of the alternate minimization steps) in favour of a regularization term involving both $\textbf{L}$ and the graph parameters via $Z(\Bar{\alpha})$. Further, the cost function used to train the neural network is the same as the objective in \eqref{eq:overalleq}, unlike \cite{chen2021time} where only one term in the objective is used to train the neural network. The proposed unrolled neural network consists of additional layers corresponding to the graph learning which are interlaced with the ones seen in \cite{chen2021time} for data inpainting.

Unrolling has also been used for graph learning in different contexts: e.g. \cite{zhang2022unrolled} proposes a distributed learning model for multi-agent collaborative setup. The goal here is to enable the agents to detect appropriate collaborators for performance gains autonomously. The graphs here denote pairwise collaborative relations which are obtained using a graph learning network; and unrolling is employed by introducing trainable attention for each model parameter at the agent. This is in contrast to our setup where the signals involved at each graph node have a time-based interpretation.

The work \cite{pu2021learning} introduces an unrolling-based technique to learn graphs with certain topological properties. The unrolling model is trained with node data and graph samples. From the graph learning perspective, this technique operates in a supervised framework. In contrast, we do not have access to graph samples, and thus operate in an unsupervised framework. Our end-to-end training is done based on how well the graph model fits the data within the inpainting process.
\vspace{-0.2cm}

\section{Results and Discussion}
In this section, we evaluate the performance of the proposed graph learning algorithm on both synthetic and real data sets.
\vspace{-0.6cm}
\subsection{Evaluation Metrics:}
\begin{itemize}
    \item \emph{Normalized error:} We evaluate the data inpainting performance by comparing the MSE error of unknown entries.

    $$\small\text{Error}= \norm{(\mathbf{1}\mathbf{1}^{\Tr}-\Psi) \circ (\hat{\mathbf{X}}-\mathbf{X})}_F \bigg/ \norm{\mathbf{1}\mathbf{1}^{\Tr}-\Psi}_1$$
    
    \item \emph{F-Score:} We use this to measure the similarity of the learnt graph with the ground truth graph (synthetic dataset only) \cite{sasaki2007truth}: higher F-Score relates to better graph learning.
 \end{itemize} 
Note that the sensing ratio is referred to as $\norm{\mathbf{1}\mathbf{1}^{\Tr}-\Psi}_1/NM$, i.e. fraction of missing entries.

\begin{figure*}[!ht]
    \centering
    \begin{subfigure}{.25\textwidth}
      \centering
      \includegraphics[width=4.2cm]{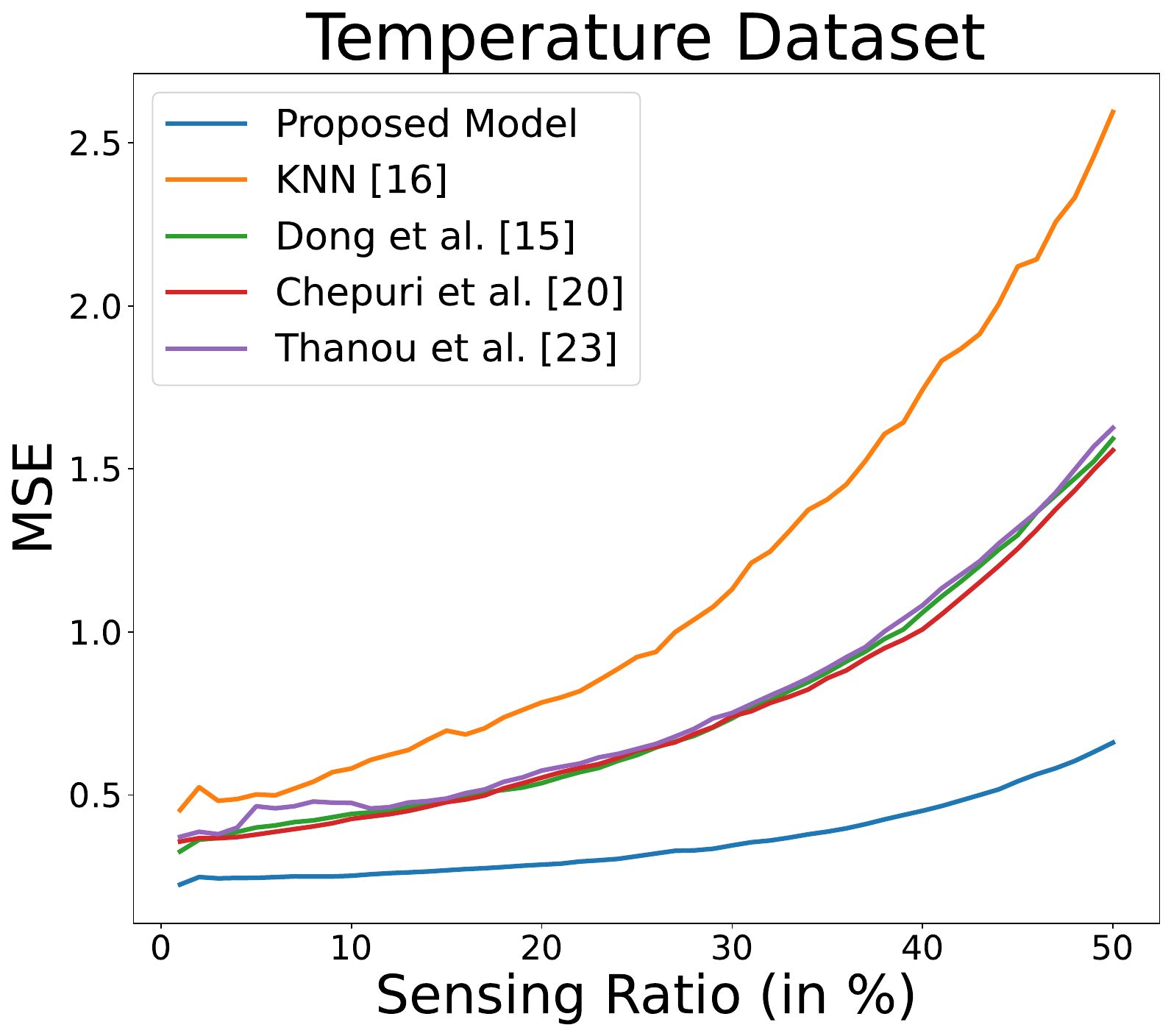}
      \caption{Temperature Dataset: MSE}
      \label{fig:MSE_brittnay}
    \end{subfigure}%
    \begin{subfigure}{.25\textwidth}
      \centering
      \includegraphics[width=4.2cm]{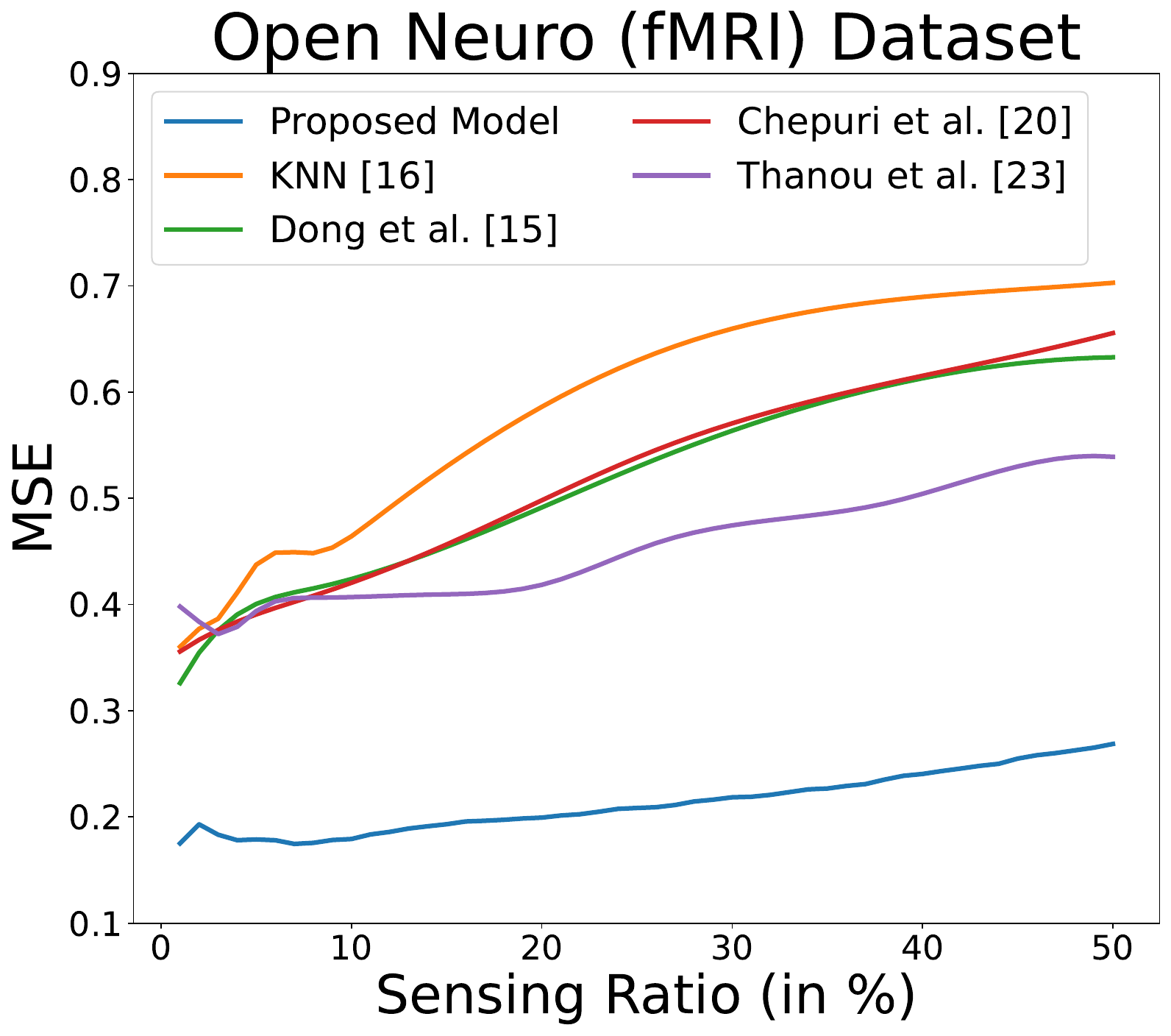}
      \caption{Open Neuro Dataset: MSE}
      \label{fig:MSE_openneuro}
    \end{subfigure}%
    \begin{subfigure}{.25\textwidth}
      \centering
      \includegraphics[width=4.2cm]{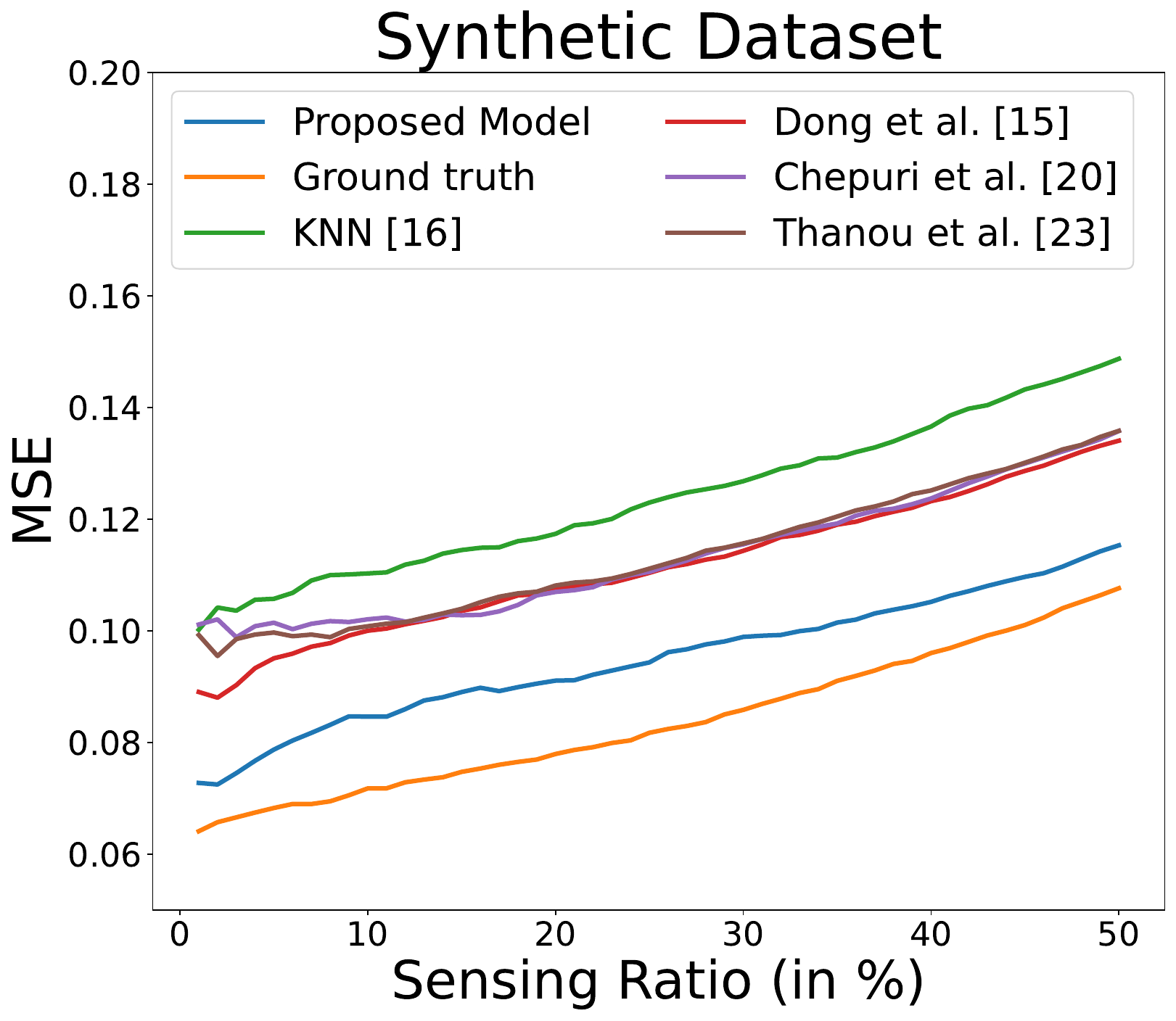}
      \caption{Synthetic Dataset: MSE}
      \label{fig:MSE_synthetic}
    \end{subfigure}%
    \begin{subfigure}{.25\textwidth}
      \centering
      \includegraphics[width=4.5cm]{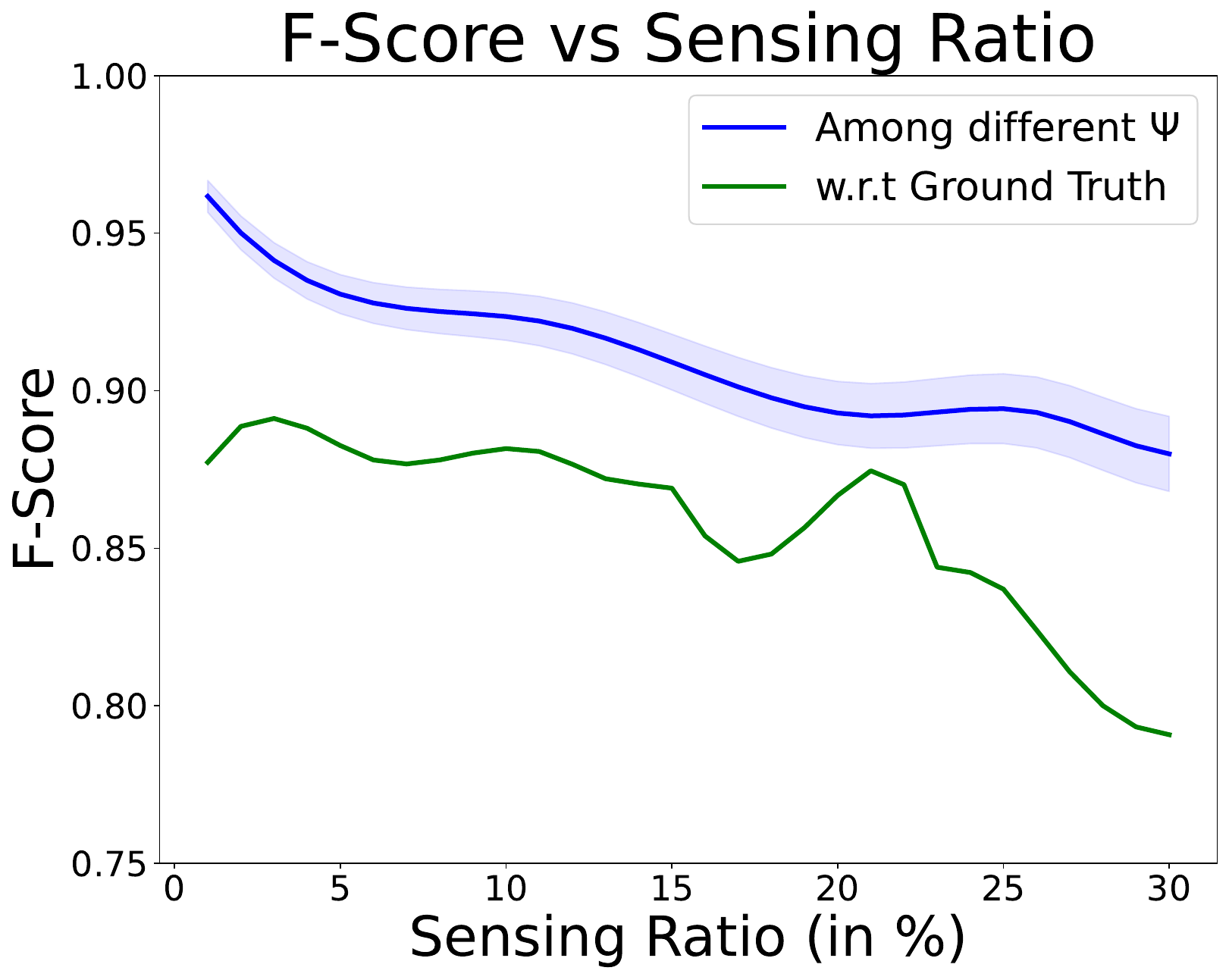}
      \caption{Synthetic Dataset: F-Score}
      \label{fig:fscore_vs_sr}
    \end{subfigure}%
    \caption{Reconstruction errors as a function of the sensing ratio on (a) Temperature dataset (b) fMRI (Open Neuro) dataset (c) Synthetic dataset (d) F-score for different sensing ratios on Synthetic dataset.}
\end{figure*}

\subsection{Competing methods}
We compare our joint graph learning and data inpainting algorithm with techniques that separately perform the two tasks. We compare with various graph learning techniques from literature including global smoothness-based learning \cite{dong2016learning, chepuri2017learning} and diffusion-based learning \cite{thanou2017learning, maretic2017graph}. We also compare with a nearest neighbour-based graph learning (each vertex is connected to its $k$ nearest neighbours in $2-$norm): as used in \cite{chen2021time}. Once the graph is learned, we artificially remove some entries and recover these missing entries using the learned graph as in \cite{chen2021time}. This process is repeated $20$ times for various mask matrices and the computed MSE is averaged.

\vspace{-0.2cm}
\subsection{Results on real datasets}
We evaluated the proposed graph learning model on two publicly available datasets. The first is the Brittany temperature dataset \cite{chepuri2017learning} (the temperature measurements collected across $32$ weather stations, with $744$ observations per weather station). The second dataset is the Open Neuro dataset \cite{bellec2017neuro} (this is fMRI data from $32$ brain regions and $152$ observations per region of interest).

%, For all these datasets, The proposed (and competing) algorithms are applied to generate the missing entries, and the MSE is computed. This process is repeated $20$ times for various mask matrices and the computed MSE is averaged.

Figure \ref{fig:MSE_brittnay} and Figure \ref{fig:MSE_openneuro} compare the performance of the proposed and competing methods on these datasets as a function of the sensing ratio. The performance gains are evident; we hypothesize that this is because the proposed algorithm can better learn the underlying graph topology, as motivated in the introduction.

\vspace{-0.2cm}
\subsection{Results on synthetic datasets}
Synthetic datasets are generated by first constructing a graph and then generating data that is a temporally smooth on the constructed graph. The (unweighted) graph is generated according Erd\H{o}s-R\'{e}nyi (ER) model \cite{erdHos1960evolution} with $N=20$ vertices and probability of edge $0.3$. On this graph, we generate time-varying data $\bf X$ as per the model in \eqref{eq:overalleq}. We generate $M=500$ time varying observations. The polynomial parameters are set as $\alpha_1=4$ and $\alpha_2=1.66$. We have verified that the proposed algorithm recovers these ground truth $\alpha$'s.

In addition to the competing methods from before, we also compare the performance of our algorithm with prior graph signal reconstruction work \cite{chen2021time} by giving the \emph{ground truth} graph as input. This is marked as a lower bound for the MSE of our algorithm.

Figure \ref{fig:MSE_synthetic} compares the data inpainting performance of the proposed and competing methods. The proposed technique shows better performance in data inpainting tasks and is quite close to the lower bound as defined above. Figure \ref{fig:fscore_vs_sr} depicts the performance of graph learning. The green plot depicting the closeness of learned graph with ground truth and blue plot showing the similarity of graphs generated with $100$ different mask matrices $\Psi$. Thus we notice that at low sensing ratios, the impact of $\Psi$ is minimal, establishing that the learnt graph is stable w.r.t where the entries in the dataset are removed.
\vspace{-0.1cm}

\begin{figure}[h!]
    \centering
    \includegraphics[width=5.5cm]{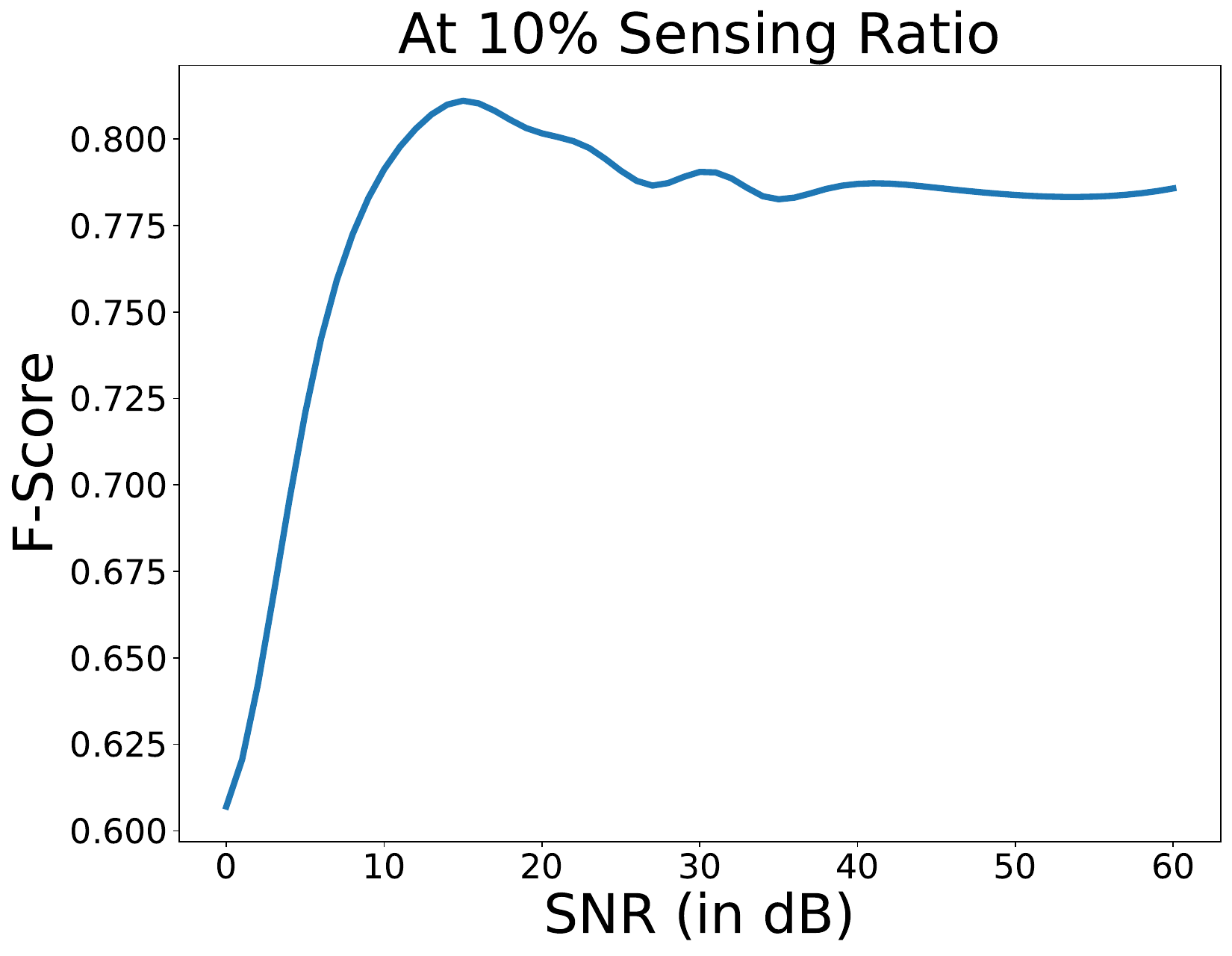}
    \caption{F-Score vs SNR on Synthetic dataset}
    \label{fig:fscore_vs_snr}
\end{figure}

To demonstrate the robustness of graph learning task, we add Gaussian noise to the (synthetic) data matrix and then apply our algorithm to learn the graph.  We then plot the F-score as a function of the SNR as seen in the Figure \ref{fig:fscore_vs_snr}.

\vspace{-0.5cm}
\section{Conclusion}
In this letter, we have proposed a new method that integrates both data-inpainting and graph-learning tasks via a closed-loop feedback system using a parameterized graph-learning model. We then unroll this algorithm into an interpretable neural network by mapping each iteration to a single layer, to learn the above-mentioned parameters. Enabling the graph learning task to be guided by the data-inpainting performance, we have demonstrated that our method outperforms other models in reconstruction error and graph learning performance. It also eliminates the dependency of knowing the underlying graph structure for the inpainting task, and hence providing a two-in-one solution. \blfootnote{The dataset and code used in this work can be downloaded from \href{https://github.com/PushkalM11/Graph-Learning-via-Unrolling.git}{here}.}

%\subsubsection{Results on Real dataset}
%We evaluated the proposed graph learning model on two different data sets. The first data set is temperature data; the temperature measurements were collected across $32$ weather stations in the French region of Brittany, and there are $744$ observations per weather station available. The second data set is the Attention Deficit Hyperactivity Disorder (ADHD) dataset. ADHD is mental pathophysiology characterized by excessive activity as a signal. The data was collected into $39$ brain regions and $350$ observations.

%The table \ref{fig:xxx} compares the difference between the reconstruction and the original time-varying graph signal on the unknown entries of the proposed algorithm with other graph learning algorithms in the literature with sensing ratio $0.3$. The learned graph from the literature is passed through the data inpainting algorithm \cite{chen2021graph}. It is evident that the proposed algorithm outperforms the existing algorithms significantly with low MSE; this is because the proposed algorithm better captures the underlying topology of the time-varying graph signals.

% \begin{figure}
%     \centering
%     \includegraphics[width=8cm]{Plots/New_Experiment/brittney_mse_vs_sr_unknown.pdf}
%     \caption{MSE vs Sensing Ratio of Unknown entries}
%     \label{fig:enter-label}
% \end{figure}

%\section{CONCLUSION AND FUTURE WORK}
%\label{sec:con}

\bibliographystyle{IEEEbib}
\bibliography{refs}

\newpage
\section{Appendix}
\label{sec:appendix}
\subsection{Estimate Missing Data (\texttt{EMD}: Section \ref{sec:updatingX})}
Since our optimization equation has different regularizers compared to the one in \cite{chen2021time}, we recompute the gradient of $f$. For the data inpainting block, we have the optimization function as follows:
\vspace{-0.2cm}
\begin{equation*}
    \small
    f(\mathbf{X}) = \norm{\Psi \circ (\mathbf{X} - \hat{\mathbf{X}})}^{2}_{F} + \lambda \; \TV_{\G}(\hat{\mathbf{X}}, \Bar{\alpha}).
\end{equation*}
Computing the gradient of $f$ with respect to \textbf{X} yields:
\begin{equation}
    \begin{aligned}
        \triangledown_{\mathbf{X}} f(\mathbf{X}) =
            2\Psi \circ (\mathbf{X} - \mathbf{\hat{\mathbf{X}}}) + 2\lambda \mathbf{L}\mathbf{X}Z(\Bar{\alpha})
    \end{aligned}
    \nonumber
\end{equation}
Further, we use equations 4a to 4d from \cite{chen2021time}, replace the gradient with the above expression and follow the exact iteration steps as mentioned in \cite{chen2021time}.

\subsection{Estimating \texorpdfstring{$\hat{\mathbf{X}}$}{X} from \cite{chen2021time}}

We use equations 4a to 4d from \cite{chen2021time} and recompute the gradient of $f$ based on our optimization problem. 
For the data inpainting block, we have the optimization function as follows:
\begin{equation*}
    f(\mathbf{X}) = \norm{\Psi \circ (\mathbf{X} - \hat{\mathbf{X}})}^{2}_{F} + \lambda \; \TV_{\G}(\hat{\mathbf{X}}, \Bar{\alpha})
\end{equation*}
Computing the gradient of $f$ with respect to \textbf{X} yields:
\begin{equation}
    \begin{split}
        \triangledown_{\mathbf{X}} f(\mathbf{X})
        &=
        \!
        \begin{multlined}[t][10cm]
            2\Psi \circ (\mathbf{X} - \mathbf{\hat{\mathbf{X}}}) + \lambda \triangledown_{\mathbf{X}} Tr(\mathbf{X}^\textsf{T} \mathbf{L} \mathbf{X}Z(\mathbf{\Bar{\alpha}}))
        \end{multlined}\\
        &=
        \!
        \begin{multlined}[t][10cm]
            2\Psi \circ (\mathbf{X} - \mathbf{\hat{\mathbf{X}}}) + \lambda \sum_{i=0}^{K} \alpha_i \triangledown_{\mathbf{X}} Tr(\mathbf{X}^\textsf{T} \mathbf{L} \mathbf{X}(\Delta \Delta^{T})^{i})
        \end{multlined}\\
        &=
        \!
        \begin{multlined}[t][10cm]
            2\Psi \circ (\mathbf{X} - \mathbf{\hat{\mathbf{X}}}) + \lambda \sum_{i=0}^{K} \alpha_i \triangledown_{\mathbf{X}} Tr(\mathbf{X}(\Delta \Delta^{T})^{i}\mathbf{X}^\textsf{T} \mathbf{L})
        \end{multlined}\\
        &=
        \!
        \begin{multlined}[t][10cm]
            2\Psi \circ (\mathbf{X} - \mathbf{\hat{\mathbf{X}}}) + \lambda \sum_{i=0}^{K} \alpha_i \mathbf{L}\mathbf{X}(\Delta \Delta^{T})^{i}
        \end{multlined}\\
        &=
        \!
        \begin{multlined}[t][10cm]
            2\Psi \circ (\mathbf{X} - \mathbf{\hat{\mathbf{X}}}) + 2\lambda \mathbf{L}\mathbf{X}Z(\Bar{\alpha})
        \end{multlined}
    \end{split}
    \nonumber
\end{equation}
The data inpainting algorithm modified with our optimization problem is as follows:
\begin{algorithm}[!htbp]
    \normalsize
    \renewcommand{\thealgorithm}{}
    \caption{\textbf{Estimating missing data from the graph}}
    \label{sec:EMD}
    \begin{algorithmic}[1]
        \Function{\texttt{EMD }}{$\mathbf{Y}$, $\mathbf{L}$, $\Bar{\alpha}$, $\lambda$, $k$}
            \State \textbf{Set:} $Z(\Bar{\alpha}) = \alpha_0 \textbf{I} + \sum_{i=1}^{k} \alpha_i (\Delta \Delta^{\Tr})^{i}$
            \State \textbf{Given:} $f^{'}(\mathbf{X}) = \Psi \circ \mathbf{X} - \mathbf{Y} + \lambda \mathbf{L} \mathbf{X} Z(\Bar{\alpha})$
            \State \textbf{Initialization:} $\mathbf{X}_0 = 0,~ \mathbf{Z}_0=-\triangledown f(\mathbf{X}_{0})$
            \For{$i \gets 0$ to $k$}
                \State $\tau \leftarrow \frac{\Tra(\triangledown f(\mathbf{X}_{i})^{T} \mathbf{Z}_i)}{\Tra(\mathbf{Y}+\triangledown f(\mathbf{X}_{i}))\mathbf{Z}_i)}$
                \State $\mathbf{X}_{i+1}  \leftarrow \mathbf{X}_{i} - \tau \mathbf{Z}$
                \State $\gamma  \leftarrow \frac{{\norm{\triangledown f(\mathbf{X}_{i+1})}^{2}_{F}}}{ {\norm{\triangledown f(\mathbf{X}_{i})}^{2}_{F}} }$
                \State  $\mathbf{Z}_{i+1}  \leftarrow -\triangledown f({\mathbf{X}_{i+1}})+\gamma \mathbf{Z}_{i}$
            \EndFor    
            \State \Return{$\mathbf{X}$}
        \EndFunction
    \end{algorithmic}
\end{algorithm}
\subsection{Semi-Norm property of graph variation term}
Here, we provide a short proof that the graph variation term $\TV_\G()$ is a semi-norm. \\
\textit{Proof:} 
Let $\Delta \Delta^{T} = \mathbf{U}\mathbf{D}\mathbf{U}^T$  where $\mathbf{U}$ is orthogonal and $\mathbf{D}$ is diagonal. Note that the diagonal entries in $\mathbf{D}$ are non-negative. We have
\begin{equation*}
    (\Delta \Delta^{T})^{i} = \mathbf{U} \mathbf{D}^{i} \mathbf{U}^{T}
\end{equation*}
Now consider our trace term 
\begin{equation}
    \begin{split}
        \TV_{\G}(\mathbf{X}, \Bar{\alpha})
        &=
        \!
        \begin{multlined}[t][10cm]
            Tr(\mathbf{X}^\textsf{T} \mathbf{L} \mathbf{X}Z(\mathbf{\Bar{\alpha}}))
        \end{multlined}\\
        &=
        \!
        \begin{multlined}[t][10cm]
            Tr(\mathbf{X}^\textsf{T} \mathbf{L} \mathbf{X}\sum_{i=0}^{k} \alpha_i \mathbf{U} \mathbf{D}^{i} \mathbf{U}^{T})
        \end{multlined}\\
        &=
        \!
        \begin{multlined}[t][10cm]
            \sum_{i=0}^{k} \alpha_i Tr((\mathbf{XU})^{T} \mathbf{L} (\mathbf{XU}) \mathbf{D}^{i})
        \end{multlined}\\
                &=
        \!
        \begin{multlined}[t][10cm]
            \sum_{i=0}^{k} \alpha_i Tr\left((\mathbf{X}\mathbf{U}\mathbf{D}^{i/2})^{T} \mathbf{L} (\mathbf{XU} \mathbf{D}^{i/2})\right)
        \end{multlined}
    \end{split}
    \nonumber
\end{equation}
Note that is a non-negative combination of non-negative terms (since the entries in $\mathbf{D}^{i}$ and $\alpha_i$ are non-negative), and hence is a semi-norm. This can also be rewritten as
\[
 \TV_{\G}(\mathbf{X}, \Bar{\alpha}) = \sum_{}\alpha_i\TV_{\G}(\mathbf{XU}\mathbf{D}^{i/2})
\]
Thus the term $\TV_{\G}()$ is a conic combination of semi-norms, and hence a semi-norm. In general, the trace of the product of two positive semidefinite matrices is non-negative (even though the product itself is not symmetric).

\subsection{Synthetic data generation}
The synthetic data generated for the experiments is as follows:
\begin{algorithm}[!htbp]
    \normalsize
    \renewcommand{\thealgorithm}{}
    \caption{\textbf{Generate synthetic data for experiments}}
    \label{alg:syn_data_gen}
    \begin{algorithmic}[1]
        \Function{\texttt{GSD }}{$N$, $\Bar{\alpha}$}
            \State Generate Erd\H{o}s-R\'{e}nyi graph $\G$ with $N$ nodes, $p=0.3$
            \State Find the eigen decomposition $(U, \Sigma)$: $U\Sigma U^\Tr = \mathbf{L_\G}$
            \State Generate $Y \sim \mathcal{N}(\mathbf{0}, \Sigma^\dagger)$, $X\leftarrow UY$ as in \cite{dong2016learning}
            \State $Z(\Bar{\alpha}) = \alpha_0 \textbf{I} + \sum_{i=1}^{k} \alpha_i (\Delta \Delta^{\Tr})^{i}$ as in \eqref{eq:Z}
        \State\Return{$XZ^{\dagger1/2}$}
        \EndFunction
    \end{algorithmic}
\end{algorithm}

\end{document}